\documentclass[fleqn,usenatbib]{mnras}

\usepackage{newtxtext,newtxmath}

\usepackage[T1]{fontenc}

\DeclareRobustCommand{\VAN}[3]{#2}
\let\VANthebibliography\thebibliography
\def\thebibliography{\DeclareRobustCommand{\VAN}[3]{##3}\VANthebibliography}

\usepackage{graphicx}	
\usepackage{amsmath}	
\usepackage{ulem}

\title[]{Synthetic disk-integrated absorption lines isolating stellar granulation for high-precision RV studies}

\author[G Frame et al.]{\parbox{\textwidth}{\Large
Ginger Frame,$^{1, 2}$\thanks{E-mail: ginger.frame@warwick.ac.uk}
Heather M. Cegla,$^{1, 2}$
Cis Lagae,$^{1, 2}$
Veronika Witzke,$^{3}$
Christopher Watson,$^{4}$
Sergiy Shelyag,$^{5}$
Vatsal Panwar,$^{1, 2}$
Michael L. Palumbo III,$^{6}$
Alexander Shapiro$^{3,7}$ 
}
\vspace{0.2cm}
\\
$^{1}$Department of Physics, University of Warwick, Gibbet Hill Road, Coventry CV4 7AL, UK\\
$^{2}$Centre for Exoplanets and Habitability, University of Warwick, Gibbet Hill Road, Coventry CV4 7AL, UK\\
$^{3}$ University of Graz, Institute of Physics, Universit\"atsplatz 5, 8010 Graz, Austria\\
$^{4}$Astrophysics Research Centre, School of Mathematics and Physics, Queen’s University Belfast, Belfast, BT7 1NN, UK \\
$^{5}$College of Science and Engineering, Flinders University, Tonsley Innovation District, 5042, South Australia, Australia.\\
$^{6}$Center for Computational Astrophysics, Flatiron Institute, 162 Fifth Avenue, New York, NY, USA\\
$^{7}$ Max-Planck-Institut f\"ur Sonnensystemforschung, Justus-von-Liebig-Weg 3, 37077 G\"ottingen, Germany\\
}

\date{Accepted 2026 February 27. Received 2026 February 26; in original form 2025 November 20}

\pubyear{\the\year{}}

\begin{document}
\label{firstpage}
\pagerange{\pageref{firstpage}--\pageref{lastpage}}
\maketitle

\begin{abstract}
We present a novel method for constructing high-accuracy, time-varying disk-integrated stellar absorption line profiles that isolate the effects of granulation alone. This framework provides an effectively unlimited supply of physically consistent training data, offering a unique opportunity to study granulation-driven velocity variability with no contamination from other stellar processes or instrumental systematics. Our interpolation scheme enables accurate profile generation at arbitrary limb angles and successfully reproduces observed disk integrated solar bisector shapes from IAG spectra. Using four \ion{Fe}{I} lines (525.0, 615.2, 617.3, and 627.1 nm), we produce 1000 model star disk-integrated realisations per line and find an isolated granulation-induced RV scatter of 0.16–0.21 m s$^{-1}$. Using our synthetic profiles and assuming infinite signal-to-noise, we find strong correlations between various line-shape metrics and convective blueshift, demonstrating that line-shape diagnostics can, in principle, trace granulation effects. Equivalent width proves the strongest diagnostic, achieving up to 60\% scatter reduction. However, the strength of all simple line shape diagnostics rapidly diminishes once photon noise is injected. Even when artificially boosting the signal to represent a spectrum containing $\sim$1000 spectral lines, the achievable improvement with these metrics remains below 10\% at typical signal-to-noise ratios. Our results highlight the need for more robust, noise-resilient diagnostics, and positions our synthetic dataset as a valuable testbed for developing and benchmarking such methods. 

\end{abstract}

\begin{keywords}
techniques: radial velocities -- Sun: granulation -- line: profiles -- hydrodynamics -- stars: solar-type -- methods: analytical
\end{keywords}



\section{Introduction}

Stellar variability signals introduce significant challenges in the interpretation of spectroscopic data, particularly in the context of exoplanet detection. These signals can arise from pressure-mode oscillations, magnetic features such as spots and faculae, granulation, and patterns of large-scale cellular flows referred to as supergranulation \citep{Sullivan2025}. Accurately isolating the radial velocity (RV) signals of small exoplanets requires effective correction for these stellar contributions. Achieving this, however, demands a deeper understanding of the distinct impact each component has on the observed spectra. A major obstacle lies in the overlapping nature of these signals, which makes them difficult to disentangle in observational data. This lack of separation limits our ability to model and correct for stellar variability, ultimately hindering progress in the detection of Earth-like exoplanets. 

Significant progress has been made in the mitigation of stellar oscillations, largely through the use of exposure times tailored to average over p-modes \citep{chaplin2019, Medina2018, gupta2022}. In the case of magnetic activity, a range of methods have been developed, including Gaussian processes (e.g: \citealt{Rajpaul2015, Jones_Gaussian, Klein2024}), neural networks (e.g: \citealt{Liang2024}), and forward models such as those in the \texttt{SOAP} series (e.g: \citealt{soap-gpu, Cristo2025}). By contrast, the mitigation of granulation and supergranulation remains the most elusive challenge. Both \citet{meunier2023} and \citet{Lakeland2024} demonstrate that unresolved granulation variability will preclude the detection of Earth-like exoplanets, even under idealized correction of other stellar noise sources.  

Granulation imprints distinct signatures on stellar absorption lines. Hot, rising granules dominate the emergent flux, producing a net convective blueshift, while the coexistence of bright upflows and dark, narrow downflows introduces characteristic line asymmetries in the form of C-shaped bisectors \citep{Gray_2005}. A key difficulty in addressing granulation signatures lies in their line dependence. Convective blueshifts and line asymmetries vary with line strength, excitation potential, formation depth, and wavelength region \citep{Dravins1981, Ramirez2009, sowmya2025, Frame2025, john2025}. To explore this complexity, \citet{grass1, grass2} developed \texttt{GRASS} (the GRanulation And Spectrum Simulator), an empirical framework for modeling granulation-induced spectral variability across 22 solar absorption lines. \citet{grass2} identified significant correlations between bisector-shape diagnostics and RV signals driven by variability, suggesting possible correction pathways. However, they also found that correlation strengths varied substantially from line to line, motivating strategies such as selectively binning similarly affected lines. These results highlight the importance of line-by-line studies for disentangling the differential impact of convection across the spectrum.  

A limitation of empirically driven approaches is the difficulty in isolating pure granulation signals. Observational spectra are invariably contaminated by residual p-mode oscillations, telluric absorption, and overlap between granulation and supergranulation. They also contain instrumental uncertainties, caused by both instrument stability and observing conditions and are limited in resolution and precision. As a result of this, the intrinsic amplitude of granulation-induced velocity fluctuations remains poorly constrained. Current estimates of the solar RV rms caused by granulation range from $\sim0.3~\mathrm{m~s^{-1}}$ \citep{Elsworth1994, dalal2023, Lakeland2024} to $\sim0.4~\mathrm{m~s^{-1}}$ \citep{Cameron2019} to $\sim 0.8~\mathrm{m~s^{-1}}$ \citep{Meunier2015}. 

\citet{cegla2013, cegla2018, cegla2019} developed a method of tiling a stellar grid with absorption lines synthesized from 3D magnetohydrodynamic (MHD) simulations produced with \texttt{MURaM} \citep{muram} and 1D radiative transfer (RT) calculations from \texttt{NICOLE} \citep{nicole1, nicole2}. The goal of this approach was to construct disk-integrated line profiles containing only the effects of granulation. Their study focused on a single spectral feature, \ion{Fe}{I} 630.2~nm, using MHD boxes representing the solar surface, with an initially imposed 200~G magnetic field.  

Building on this, \citet{Frame2025} (hereafter F25) used updated hydrodynamic (HD) \texttt{MURaM} simulations \citep{witzke2024} also for the solar case but with no magnetic field, together with \texttt{MPS-ATLAS} \citep{mps-atlas} RT calculations to extend the analysis to four \ion{Fe}{I} lines. Crucially, F25 introduced an updated parameterisation that explicitly removes the artificial p-mode oscillations present in HD simulations; this is an essential step given that the p-mode filtering in \citet{cegla2013} was only adequate for the more strongly damped oscillations in MHD boxes. This refinement ensures a clean isolation of granulation signatures, overcoming a key limitation of observational data in which granulation is blended with p-modes, tellurics, and supergranulation.

F25 also investigated the center-to-limb dependence of granulation, showing that bisector shape, convective blueshift, and RV rms each exhibit non-trivial, line-dependent relationships with limb angle. Similar behavior was reported by \citet{cegla2018}, motivating the use of fine $2^\circ$ sampling when tiling the stellar surface in subsequent simulations \citep{cegla2019}. Such sampling is fully adequate for capturing granulation signals, but is extremely computationally expensive, particularity for multi-line analyses. 

In this work, we adopt a continuous sampling approach using an interpolation scheme implemented in \texttt{DISCO} (Disk Integrated Stellar COnvection)\footnote{\url{https://github.com/ginger-frame/DISCO}}, a publicly available set of Python scripts for generating and analysing disk-integrated stellar absorption line profiles. \texttt{DISCO} is capable of computing time-varying line profiles at arbitrary limb angles without the need for additional radiative transfer calculations. This approach reduces computational expense while maintaining the accuracy of the disk-integration process by enabling continuous and physically consistent coverage of the stellar disk. 

It is worth briefly contrasting this framework with other tiled stellar surface models used in the literature. Statistical approaches such as that of \citet{Meunier2015} describe granulation through prescribed distributions of granule properties (e.g. size, velocity field, and lifetime), rather than through explicit synthesis of spectral line profiles at the tile level. Forward models such as \texttt{SOAP} are primarily designed to capture the effects of magnetic active regions and do not currently model time-resolved granulation-driven line profile variability, while empirical tools such as \texttt{GRASS} require observed time series as input and are therefore not predictive in the absence of suitable data. The approach adopted here is closely related to that of \citet{cegla2019}, but differs in our use of a line-profile–based parameterisation of unresolved granulation combined with continuous limb-angle interpolation, as discussed earlier. 

In this work, we use this new interpolation-based tiling scheme to construct time-varying, disk-integrated stellar absorption line profiles that contain only the effects of granulation. This allows us to isolate the intrinsic velocity variability caused by granulation and to assess, under idealised conditions, the extent to which it can be mitigated using common spectral diagnostics. We test both bisector asymmetry and full-line shape metrics across four \ion{Fe}{I} lines to determine their ability to trace and correct convective variability. By subsequently introducing realistic instrumental resolution and injecting various levels of photon noise, we evaluate the practical limits of these diagnostics and quantify the residual RV scatter that would remain in high-precision observations. Through this approach, we aim to establish both the theoretical lower bound on granulation-induced velocity noise and the observational requirements needed to overcome it. 

A summary of the data used in this work is provided in Section \ref{sec:data}. For a more in-depth description, as well as a full explanation of the parameterisation process, we refer the reader to F25. Section \ref{sec:lps} outlines our interpolation scheme for generating profiles at arbitrary limb angles. Section \ref{sec:sg} describes our stellar grid set up. Section \ref{sec:DI} explains the calculation of our disk-integrated profiles using \texttt{DISCO}. Section \ref{sec:corr} explores techniques for the mitigation of granulation induced RVs under idealised circumstances. Section \ref{sec:noise} investigates the impact of signal-to-noise ratio (SNR) on our ability to trace granulation signatures in stellar spectra.  

\section{Data}
\label{sec:data}

\begin{table}
\centering
\caption{Summary of simulation and line synthesis setup.}
\begin{tabular}{ll}
\hline
\textbf{Property} & \textbf{Value / Description} \\
\hline
Simulation code & \texttt{MURaM} \citep{muram, witzke2024} \\
Simulation duration & $\sim$1 hour \\
Cadence & 30 seconds \\
Horizontal extent & $9 \times 9$~Mm$^2$ \\
Horizontal resolution & $512 \times 512$ pixels \\
Vertical extent & 5~Mm$^{*}$ \\
Vertical resolution & 500 pixels \\
\hline
Line synthesis code & \texttt{MPS-ATLAS} \citep{mps-atlas} \\
Spectral resolution & $R = 2{,}000{,}000$ \\
\hline
\end{tabular}

$^{*}$ 1~Mm of atmosphere above the photosphere and 4~Mm of sub-photosphere and convection zone of a solar type star.

\label{tab:simulation_summary}
\end{table}

\begin{table*}
    \centering
    \begin{tabular}{l c c c c c c c c}
    \hline
    Element + ionisation & Air wavelength (nm) & log(\textit{gf}) & Land\'e factor& $\gamma_{\text{rad}} (s^{-1}) $ & $\gamma_{\text{Stark} }(s^{-1})$ & $\gamma_{\text{VdW} }(s^{-1})$ & Excitation potential (eV) \\
    \hline
    \ion{Fe}{I}  & 525.021 & -4.938$^{\dagger}$  & 3.00    & 3.32  & -6.28 & -7.82$^*$ & 0.121 \\
    \ion{Fe}{I}  & 615.162 & -3.299$^{\dagger}$ & 1.84 & 8.29 & -6.16 & -7.70$^*$ & 2.176 \\
    \ion{Fe}{I}  & 617.333 & -2.880$^{\dagger}$  &  2.50  & 8.31  & -6.16 & -7.69$^*$ & 2.223 \\
    \ion{Fe}{I} & 627.128 & -2.701$^{\ddagger}$ &  1.49 & 8.23 & -5.41 & -7.28$^*$ & 3.332 \\
    \hline
    \end{tabular}
    \caption{Atomic parameters for the lines selected for this work. $\gamma_{\text{rad}}$, $\gamma_{\text{Stark}}$ and $\gamma_{\text{VdW}}$ refer to the radiative, Stark and Van der Waals damping parameters. All data is sourced from \protect\cite{K14}, unless marked otherwise. \\[0.5ex]
    \textbf{Key:} $^\dagger$ Source: \protect\cite{FMW}; $^\ddagger$ Source: \protect\cite{BK}; $^*$ Source: \protect\cite{BPM}.}
    \label{tab:atomic_data}
\end{table*}

All absorption line profiles in this work were synthesised from a time series of 3D hydrodynamic simulation boxes. To produce line profiles at different positions on the stellar disk and account for changes in limb darkening and convective blueshift, the hydrodynamic quantities of the 3D cubes were tilted and interpolated onto rays parallel to the desired limb angle, prior to line synthesis (see \citealt{Smitha2025} for a description of this process). For details on our data and line synthesis method, see section 2 in F25. A summary of the simulation properties is provided in Table \ref{tab:simulation_summary}.

Line synthesis for the full time series was performed for the following lines: \ion{Fe}{I} 525.0~nm, \ion{Fe}{I} 615.2~nm, \ion{Fe}{I} 617.3~nm, \ion{Fe}{I} 627.1~nm.

For the atomic properties of these lines, see Table \ref{tab:atomic_data}.

\section{Producing profiles at arbitrary limb angles}
\label{sec:lps}


In F25, we demonstrated that absorption line profiles at different limb angles can be parameterized into three components: granular tops (GTs), outer granular regions (OGRs), and intergranular lanes (IgLs). By averaging these component profiles over time and space, we effectively remove p-mode oscillations while preserving the granulation signal in the component filling factors. This yields a valuable dataset consisting of time-averaged component profiles and their corresponding filling factor distributions. By sampling new filling factors from these distributions and combining them with the component profiles, we can synthesize additional accurate line profiles. In this work, we show that we are able to construct highly accurate line profiles at any given limb angle for unlimited instances in time. 

We analysed the filling factor distributions from F25 and found a strong linear correlation between OGR and IgL filling factors, but no significant correlation involving GT filling factors. Based on this, we produce new synthetic profiles by independently sampling GT filling factors from their distribution, and separately sampling from the distribution of the OGR-to-IgL filling factor ratio. 

A random instance of a line profile at a given limb angle $\theta$ (see Section \ref{ss: limb_angle} for $\theta$ definition) is then given by:

\begin{equation}
\begin{split}
I(\theta) &= I_{\mathrm{GT}}(\theta)\epsilon_{1}
+ I_{\mathrm{OGR}}(\theta)\frac{\epsilon_{2}(1-\epsilon_{1})}{1+\epsilon_{2}}
+ I_{\mathrm{IgL}}(\theta)~\frac{(1-\epsilon_{1})}{1+\epsilon_{2}} \\\\
\epsilon_{1} &\sim \mathcal{D}_{\mathrm{GT}}(\theta) \\
\epsilon_{2} &\sim \mathcal{D}_{\mathrm{OGR}/\mathrm{IgL}}(\theta)\\
\end{split}
\label{eq:I}
\end{equation}
where each $I_{C}(\theta)$ ($C \in \{\mathrm{GT}, \mathrm{OGR}, \mathrm{IgL}\}$) is the average component  profile and each $\mathcal{D}(\theta)$ is the probability distribution from which the relevant filling factor $\epsilon$ is pulled. 

To be able to generate $I(\theta)$ at any given $\theta$, we must be able to interpolate on $\theta$ each $I_{C}(\theta)$ and $\mathcal{D}(\theta)$. To do this we first define the optimum set of training limb angles, based upon the trade-off between computation time and minimisation of root mean square error (rmse) in the disk-integrated profile. We find the optimal training set consists of 16 angles (in degrees): 0, 2, 6, 9, 13, 17, 20, 23, 27, 35, 50, 60, 68, 81, 82, and 83. For a description of the method used to determine these angles, see Appendix \ref{sec:A1}. 

In F25, calculations were performed from $\mu$ = 1.0 to $\mu$ = 0.2 in steps of 0.1, where $\mu = \cos(\theta)$. Other than $\mu$ = 1.0 ($\theta$ = 0$^{\circ}$) and $\mu$ = 0.5 ($\theta$ = 60$^{\circ}$), the results in F25 are not included in the training set for this work. We therefore use these results as our validation set to check the interpolation is working as expected. 

\subsection{Interpolating average component profiles}
\label{ss: PCA}

\begin{figure*}
    \centering
    \includegraphics[width=\textwidth]{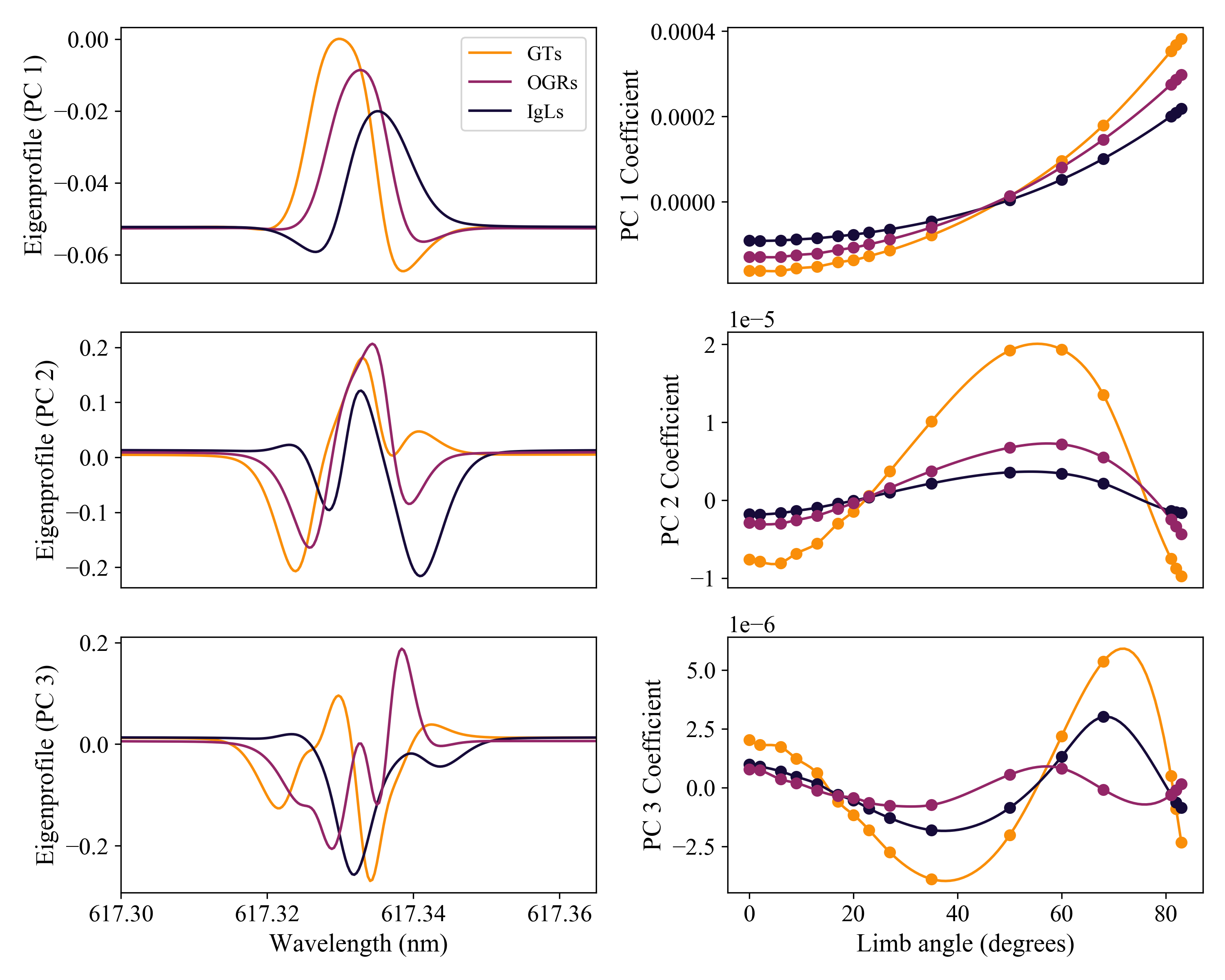}
    \caption{First three eigenprofiles (principal components) of each component profile (left column) and the associated coefficients and their variation with limb angle (right column) for \ion{Fe}{I}~617.3nm. The circular points show results for the training set of limb angles from which the interpolation is based on.}
    \label{fig:pca}
\end{figure*}

To interpolate the component line profiles across different stellar limb angles, we employ Principal Component Analysis (PCA). This allows us to reduce the dimensionality of the line profiles and interpolate their variation smoothly with angle.

PCA identifies a set of orthogonal basis functions called principal components (PCs), or eigenprofiles, that capture the main variations in the profiles across different limb angles. Each original profile can then be approximated as a weighted sum of these components, plus a mean profile that represents the average shape across all angles. The weights (also called PCA coefficients or scores) vary with limb angle and describe how much each principal component contributes to a given profile. Each coefficient \(a_k(\theta)\) is interpolated as a smooth function of \(\theta\) using cubic splines. 

Figure \ref{fig:pca} shows the first three PCs for the three average components in the \ion{Fe}{I} 617~nm line, along with the variation of the coefficients with $\theta$. In practice, we retain seven PCs for each component, which we find provides a good balance between capturing the relevant variability in the data and avoiding overfitting. See Appendix \ref{sec:A1} for more details. 

At an arbitrary $\theta$, each component profile is constructed as:

\begin{equation}
I_{C}(\theta) \approx \bar{I_{C}} + \sum_{k=1}^{7} \tilde{a}_k(\theta)\, \phi_k.
\end{equation}

\noindent where $\bar{I_C}$ is representative of the mean profile over all $\theta$, $\tilde{a}_k(\theta)$ is the spline-interpolated coefficient for component \(k\), and $\phi_k$ is the PC. 

\subsection{Characterising and interpolating filling factor distributions}
\label{ss:ff}

\begin{figure*}
    \centering
    \includegraphics[width = \textwidth]{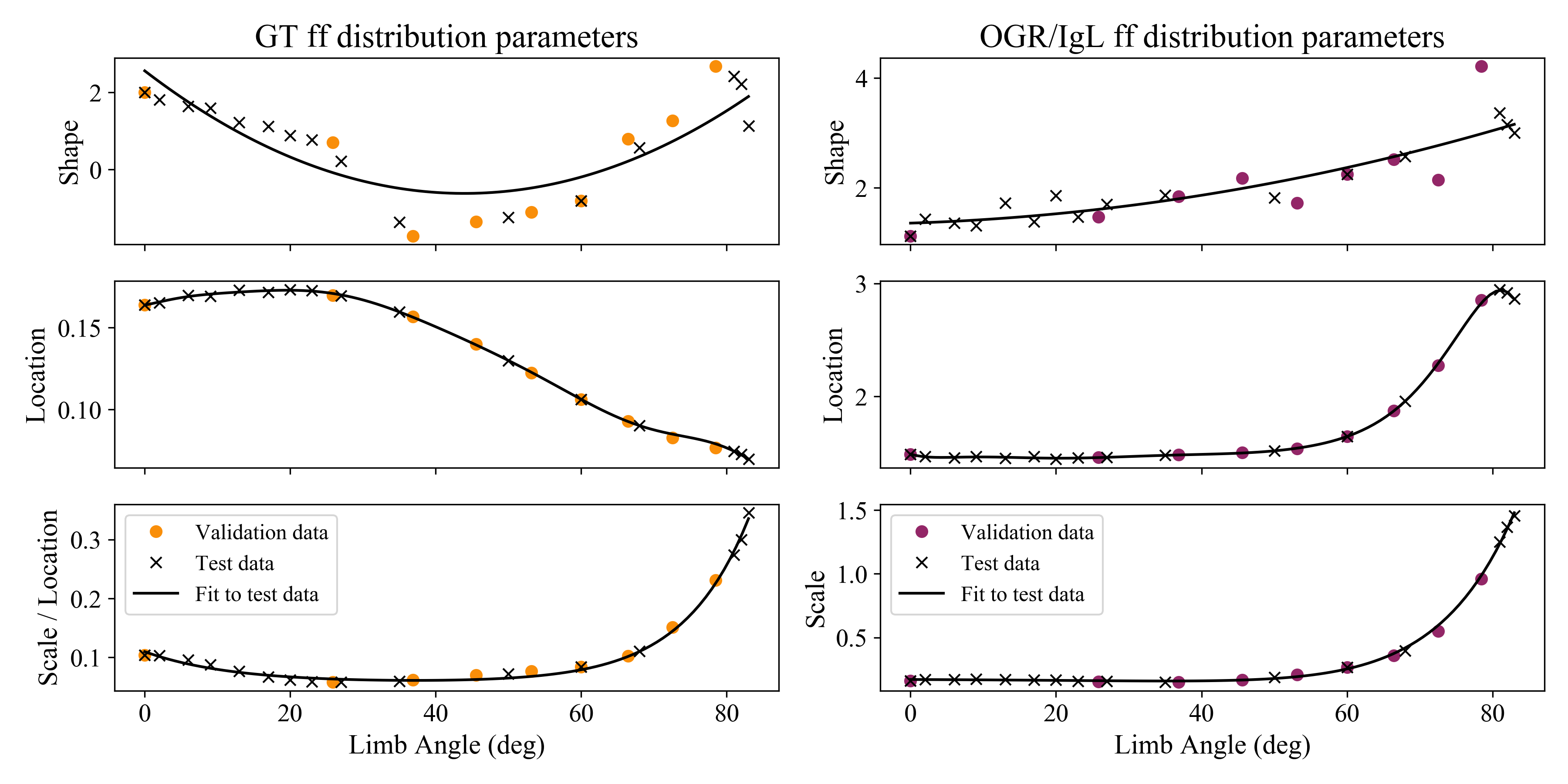}
    \caption{Model fits to the parameters required to build skewed Gaussian distributions for both the GT filling factors (left column) and the ratio OGR/IgL filling factors (right column) for \ion{Fe}{I}~617.3nm. The black crosses show the best fit parameters to the test-set distributions, and the black line is the model fit to these data. The colored circles show best fit parameters to the validation-set distributions, these values are not included in the fitting. Note that for the GT filling factor distribution the scale parameter is normalised by the location parameter. This is not the case for the OGR/IgL filling factor distribution.}
    \label{fig:ff_fits}
\end{figure*}

\begin{figure*}
    \centering
    \includegraphics[width = \textwidth]{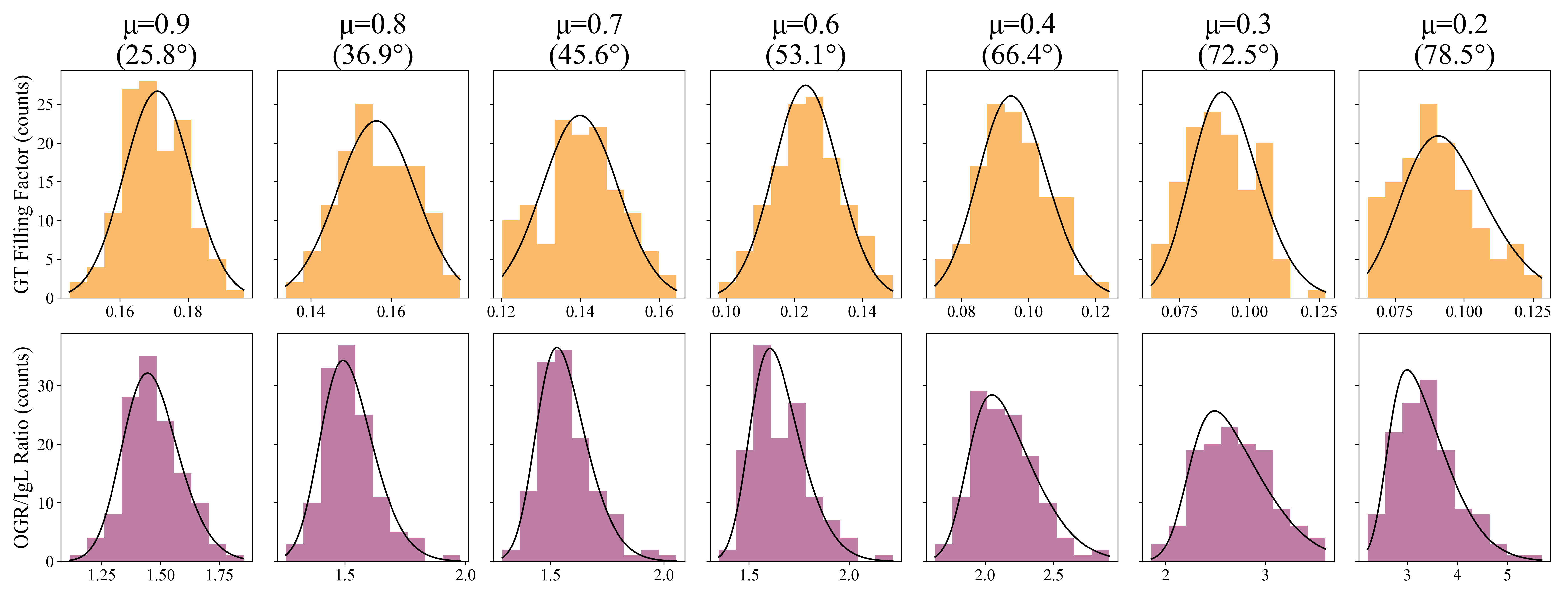}
    \caption{Histograms of GT filling factors and OGR/IgL ratio filling factors for the validation set of angles for \ion{Fe}{I}~617.3nm. Overplotted in black are the predicted skewed Gaussian distributions using parameters derived through the interpolation procedure using a seperate set of training angles. These Gaussians are not fit to the data shown and serve as a test of whether distributions at the validation angles are being correctly captured.}
    \label{fig:gauss_fits}
\end{figure*}

In F25, we established that granulation-induced velocity and line-shape variations are preserved in our time series of filling factors. It is therefore crucial to accurately maintain the underlying statistical distributions when drawing new samples.

Following the approach of \citet{cegla2019}, one option of characterising the filling factor distributions is to fit generalised logistic functions to the cumulative distribution functions (CDFs). While this method performs well when fitting each training angle independently, we find that the resulting fit parameters do not vary smoothly with limb angle, making interpolation problematic.

To address this, we instead model the distribution of filling factors at each training angle using a skewed Gaussian distribution, characterised by three parameters: shape $\alpha$ (controlling skewness), location $\zeta$ (analogous to the mean), and scale $\omega$ (analogous to the standard deviation). Although this model is simpler than the generalised logistic function, it proves sufficiently flexible for our application. Specifically, a Kolmogorov–Smirnov (KS) test between samples drawn from the fitted skewed Gaussian and the original training data consistently passes at all limb angles, confirming the model's adequacy.

All choices regarding the interpolation of the skewed Gaussian parameters are guided by the success of the KS test applied to the validation set of limb angles. It is important to note that this validation set is entirely excluded from the fitting procedure and serves solely to assess generalization. Specifically, for each validation angle, we draw 129 random samples from the interpolated distribution, matching the size of the validation set, and repeat this sampling process 1000 times. The KS test is performed on each sample set, guiding interpolation choices to ensure statistical consistency at unseen angles. The KS test returns two key metrics: the D statistic, which quantifies the maximum difference between the cumulative distributions (with 0 indicating a perfect match and values closer to 1 indicating large discrepancies), and the p-value, which estimates the likelihood of observing such a difference under the assumption that both samples are drawn from the same distribution. A p-value greater than 0.05 typically indicates that the null hypothesis (that the distributions match) cannot be rejected. When we refer to the KS test `passing', we mean that the p-value is over 0.05 and the difference between the distributions are not statistically significant. 

For both \(\mathcal{D}_{\mathrm{GT}}\) and \(\mathcal{D}_{\mathrm{OGR/IgL}}\), we find that the shape parameter \(\alpha\) must be interpolated over $\theta$ using a second-order polynomial. Although this provides only a modest fit to the training data, increasing the polynomial order leads to overfitting. While the training RMSE continues to decrease, the RMSE evaluated on the validation set increases. Omitting the skewness term altogether significantly degrades the results. To mitigate the impact of this suboptimal fit to the shape parameter, we fix \(\alpha\) to its interpolated value and then refit the location and scale parameters accordingly.

The location parameter \(\zeta\) is well captured by a 10th-order polynomial in both distributions. For the scale parameter \(\omega\), the optimal interpolation method differs between the two cases. In \(\mathcal{D}_{\mathrm{GT}}\), we achieve the best results by normalizing \(\omega\) by the corresponding \(\zeta\), and fitting the normalized scale using a sum of two exponential functions. In contrast, for \(\mathcal{D}_{\mathrm{OGR/IgL}}\), the best fit is obtained by applying the sum of exponentials directly to the unnormalized scale parameter.

Figure~\ref{fig:ff_fits} shows the fitted parameters and their interpolation curves for both distributions. The colored points correspond to fits from the validation sample and are not used in constructing the interpolation, serving as an independent check of the model's performance.

For \(\mathcal{D}_{\mathrm{OGR/IgL}}\), the interpolated distributions pass the KS test in more than 97\% of trials at all validation angles. The results for \(\mathcal{D}_{\mathrm{GT}}\) are similarly robust, with success rates exceeding 97\% except at \(\mu = 0.3\) (\(\theta = 72.54^\circ\)) and \(\mu = 0.2\) (\(\theta = 78.46^\circ\)), where the pass rate decreases to approximately 70\%. However, at such extreme limb angles, the GT component contributes minimally to the overall disk-integrated signal with filling factors < 0.1, therefore this decline in fit quality is not expected to significantly impact the final disk integrated results. Note also that tiles towards the limb are heavily limb-darkened and fore-shortened, and so their accuracy holds less weight than those nearer disk center when it comes to preserving disk-integrated accuracy.  

Figure~\ref{fig:gauss_fits} shows the predicted skewed Gaussian distributions for the validation set of angles based on the interpolation procedure. The distributions are overplotted on histograms of the actual distributions within the validation set. The predicted distributions appear to effectively contain the real distributions, suggesting the interpolation of parameters has been successful. Some degradation of effectiveness is evident towards the limb, as discussed above. For an alternative test of our ability to correctly capture the filling factor distributions, see Appendix \ref{sec:B1} for Quantile-Quantile (Q-Q) plots.

\subsection{Validation of interpolation}
\label{ss:validation}

\begin{figure*}
    \centering
    \includegraphics[width = \textwidth]{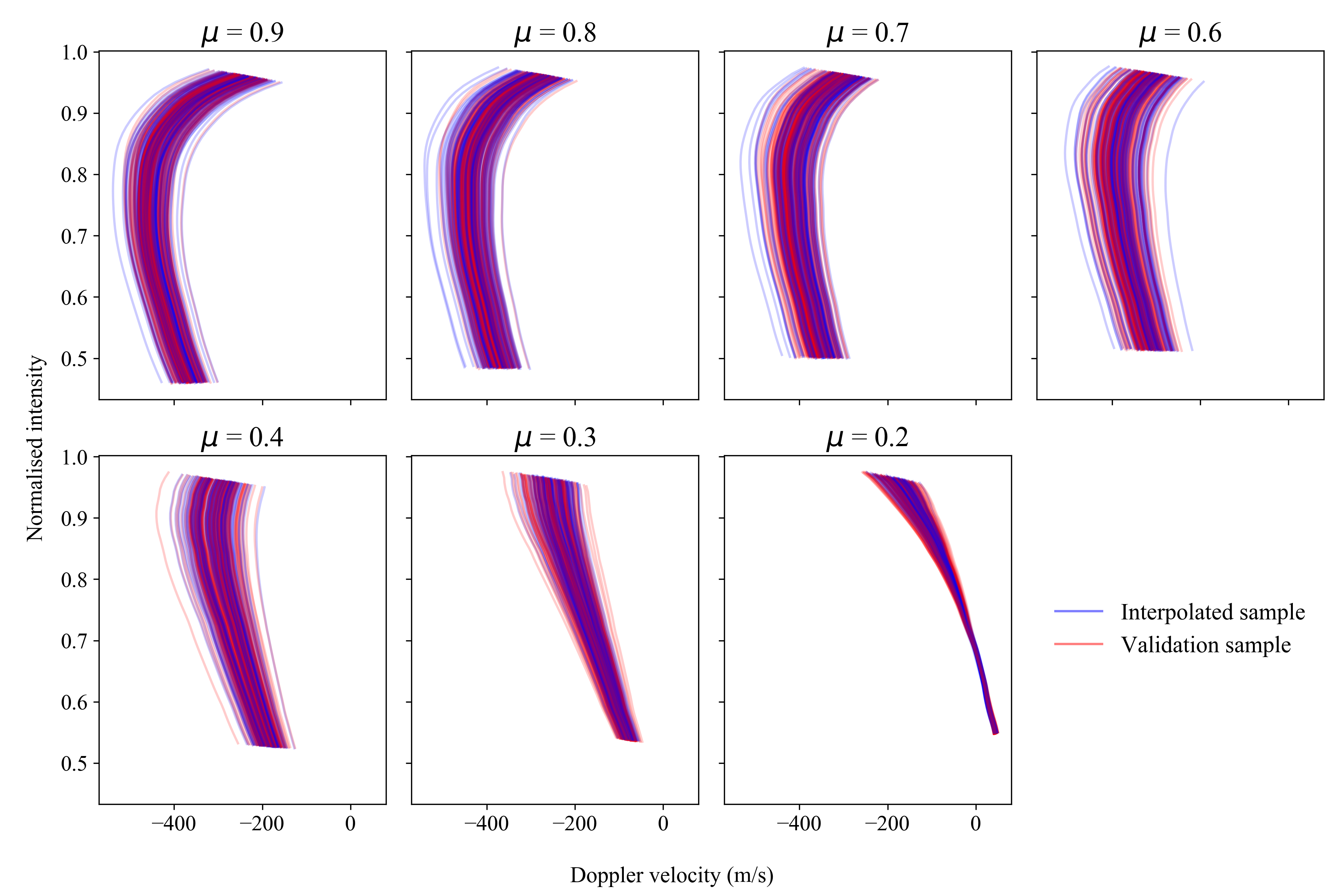}
    \caption{\ion{Fe}{617.3}~nm bisectors from an example sample pulled from the interpolation method at each validation limb angle. Bisectors from the validation time series are overplotted. The interpolation and validation sample at each angle are of equal size. In an attempt to mitigate overlap, we alternate plotting interpolated and validation bisectors. The intensity values have been normalised in each case by the maximum intensity value in the validation sample at the relevant angle.}
    \label{fig:validation}
\end{figure*}

Using our interpolated components from Section \ref{ss: PCA} and our interpolated distributions from Section \ref{ss:ff} we construct a sample of profiles for each validation angle using equation \ref{eq:I}, matching the size of the validation time series (129 instances). Figure \ref{fig:validation} shows the bisectors from these samples, alongside the bisectors from the validation data for \ion{Fe}{I}~617~nm. This plot serves as a visual demonstration that general bisector shapes and velocities are being preserved in the interpolated samples. 

To assess how accurately the velocity distributions are preserved, we use the peak of a cross-correlation function (CCF) as a measure of radial velocity. We calculate the CCF using the relevant time-averaged validation profile as a template, and measure the peak with a second order polynomial fit to 3 data points. For each validation angle, we perform KS tests comparing the validation velocities to 1000 independently generated sample velocity distributions of matching size. The results are summarized in Table~\ref{tab:ks_results}. Note that for a two-sample KS test where each sample has a size 129, the critical D value lies at $\sim$ 0.169, and so any D value falling under this provides evidence for matching distributions. Across all validation angles, the majority of tests return p > 0.05, with pass rates in the range 82-96\%. The pass rates are not expected to be 100\%: even two draws from the same underlying distribution will fail a KS test a non-negligible fraction of the time when the sample size is modest. The observed pass rates include expected statistical fluctuations associated with repeated finite sampling, and do not indicate systematic discrepancies in the interpolated distributions. Overall, we find that the interpolation preserves the underlying velocity statistics to within the limits imposed by finite-sample effects.

\section{Producing the stellar grid}
\label{sec:sg}

\begin{figure*}
    \centering
    \includegraphics[width = \textwidth]{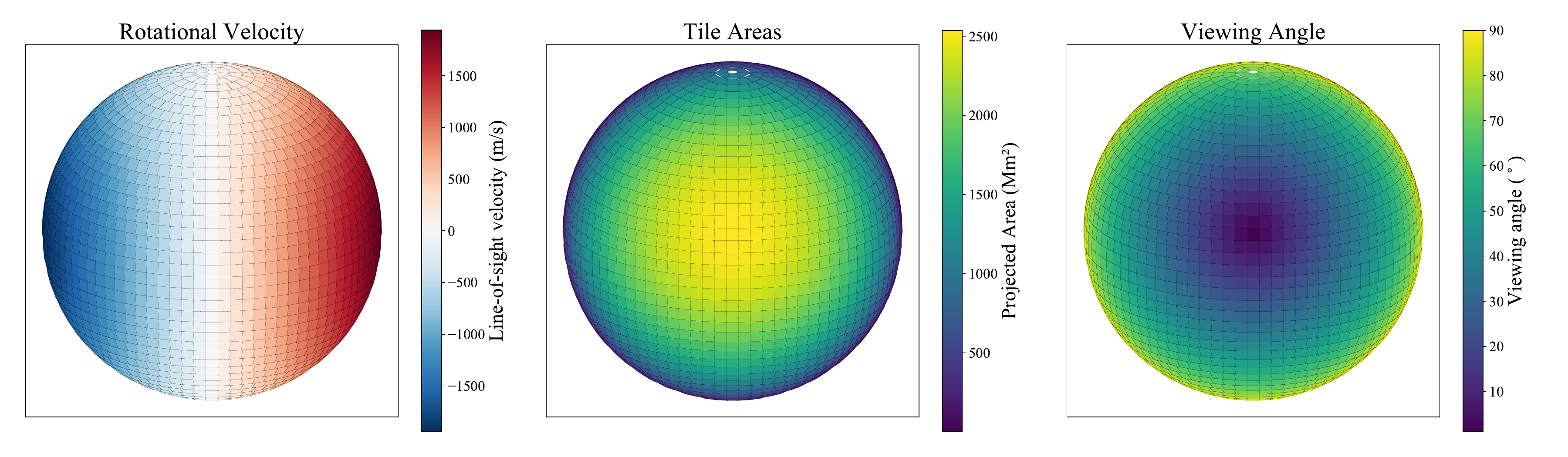}
    \caption{An example stellar grid with enlarged tiles and slight inclination for demonstrative purposes. The left hand plot shows the line of sight stellar rotational velocity at each tile, including differential rotation. The middle plot shows the projected area of each tile. Note that these are enlarged tiles, hence the inflated area values. The rightmost plot shows the limb angle at each tile.}
    \label{fig:grid}
\end{figure*}

We follow the methodology in \cite{cegla2019} for the production of our stellar grid. See also \cite{vogt1987} and \cite{Piskunov2002} for similar procedures. The python scripts used to create the grid are available for use, stored in the \texttt{DISCO} repository. 

We construct a sphere of radius $R_{\odot}$ with tiles of size $9 \times 9~\mathrm{Mm^2}$ to match the surface area of our \texttt{MURaM} simulation boxes. We then project the visible half of the sphere onto 2D space. For each visible tile we calculate and store the line of sight velocity, projected area and limb angle. The calculations and uses for each of these parameters are described in the subsections below. Figure \ref{fig:grid} depicts an example grid color-coded by each parameter. The tiles in this figure have been made larger for demonstrative purposes, and the star has been inclined by 70$^\circ$. The true setup used in this study contains $\sim$37,000 visible tiles and has an inclination of 90$^\circ$ (face on).

\begin{table}
\centering
\caption{Results of a 2-sample KS test performed on velocity values calculated using the validation set and a randomly generated sample using the interpolation procedure. For each limb angle ($\theta$, also displayed as $\mu = cos(\theta)$), the test is run 1000 times using unique occurrences of the generated samples (see Section \ref{ss:validation}).}
\begin{tabular}{cccccc}
\hline
$\mu$ & $\theta$ (deg) & Passed (\%) & Avg.\ D & Avg.\ p-value \\
\hline
0.9 & 25.8  & 92.1 & 0.1222 & 0.3593 \\
0.8 & 36.9  & 90.5 & 0.1191 & 0.4057 \\
0.7 & 45.6  & 82.2 & 0.1313 & 0.3175 \\
0.6 & 53.1  & 87.8 & 0.1268 & 0.3386 \\
0.4 & 66.4  & 96.0 & 0.1161 & 0.4042 \\
0.3 & 72.5  & 82.7 & 0.1390 & 0.2308 \\
0.2 & 78.5  & 90.6 & 0.1349 & 0.2385 \\
\hline
\end{tabular}
\label{tab:ks_results}
\end{table}

\subsection{Line of sight velocity}
\label{ss:losvel}

We use the solar differential rotation model from \cite{Snodgrass1990} to calculate the angular rotational velocity $\omega$ as a function of latitude $\phi$:


\begin{equation}
\begin{split}
    \omega(\phi) &= A + B~\sin^2(\phi) + C~\sin^4(\phi) \\
    A &= 2.773 \times 10^{-6}~\text{rad s}^{-1} \\
    B &= -4.84 \times 10^{-7}~\text{rad s}^{-1} \\
    C &= -3.61 \times 10^{-7}~\text{rad s}^{-1}
\end{split}
\end{equation}

\noindent where we have subtracted Earth's orbital angular velocity from the constant coefficient $A$ to convert from sidereal to synodic velocity.

The tangential velocity $v_{t}(\phi)$ at each latitude is given by $R_{\odot}\omega(\phi)\cos(\phi)$. The velocity value that we want to store is the tangential velocity projected along the line of sight to the observer. This is calculated as follows:

\begin{equation}
    v_{los}(\phi, \lambda) = v_{t}(\phi)\cdot\sin(\lambda)~\sin(i)
\end{equation}

\noindent where $\lambda$ is the longitude and $i$ is the stellar inclination.

\subsection{Projected area}
\label{ss:proj_area}

To cover the stellar surface with approximately square tiles, we sample the surface at a discrete set of latitudes $\{\phi_j\}$ uniformly spaced between $-90^\circ$ and $+90^\circ$, with the spacing chosen such that the arc length of each latitudinal slice is approximately equal to the simulation box size (9~Mm). At each discrete latitude $\phi_j$, the number of tiles along the corresponding parallel is given by:

\begin{equation}
N_{\phi_{j}} = \frac{2\pi R_{\odot} \cos\phi_{j}}{9\mathrm{Mm}}.
\end{equation}

where $N_{\phi_j}$ is rounded to the nearest integer. Longitude is sampled at discrete points $\{\lambda_k\}$ along each parallel such that the horizontal extent of each tile is also approximately 9~Mm. The surface area of an individual tile is then obtained by dividing the area of the latitudinal slice at $\phi_j$ by $N_{\phi_j}$ along that parallel.

The projected area refers to the portion of the tile visible to an observer when the 3D sphere is projected onto the 2D sky plane. Cartesian coordinates for each grid point are computed as:

\begin{equation}
    \begin{split}
        x = R_{\odot}~\cos(\phi)~\sin(\lambda)\\
        y = R_{\odot}[\sin(\phi)~\sin(i)~-~\cos(\phi)~\cos(\lambda)~\cos(i)]\\
        z = R_{\odot}[\cos(\phi)~\cos(\lambda)~\sin(i)~+~\sin(\phi)\cos(i)]
    \end{split}
    \label{eq:coords}
\end{equation}

\noindent where the $xy$-plane is the plane of the sky, and the $z$-axis points towards the observer. The coordinate system is defined such that when $i = 90^\circ$, the point at $(\phi, \lambda) = (0, 0)$ maps to $(x, y, z) = (0, 0, R_{\odot})$.

The projected area $A_{p}$ of a tile is then calculated as:

\begin{equation}
A_{p} = V~A_{t} \sqrt{1 -  \left({\frac{x_c}{R_{\odot}}}\right)^{2} -  \left({\frac{y_c}{R_{\odot}}}\right)^{2}}
\end{equation}

\noindent where $A_{t}$ is the surface area of the tile, and $(x_c, y_c)$ are the coordinates of its center. The factor $V$ is the visibility factor, representing the fraction of the tile that lies on the visible hemisphere. Points with $z > 0$ are considered visible. To determine $V$, we compute $z$ for all four corners of the tile:

\begin{itemize}
    \item If all corners are visible, $V = 1$
    \item If none are visible, $V = 0$
    \item If partially visible, we subdivide the tile into 100 smaller sub-tiles and compute the fraction with visible centers. This fraction becomes $V$.
\end{itemize}

Enforcing approximately square tiles on a spherical surface introduces small geometric discontinuities near the rotational poles, where the number of tiles per latitudinal slice decreases. These discontinuities are visible upon inspection in Figure \ref{fig:grid}, they are however visually exaggerated due to the tiles having been significantly enlarged. Because tile areas are defined analytically from each latitudinal slice, the total surface area per slice is conserved by construction, even if gaps or overlaps between tiles occur. For viewing geometries close to edge-on ($i \simeq 90^\circ$), the impact of the discontinuities are further suppressed by geometric foreshortening and limb darkening. While their contribution may increase for non-face-on configurations, the large number of tiles and the conservation of total surface area should ensure that their effect on disk-integrated quantities remains negligible.

\subsection{Limb angle}

\label{ss: limb_angle}

The limb angle $\theta$ refers to the angle between the surface normal and the line of sight: 

\begin{equation}
    \theta = \cos^{-1} \left( \frac{z}{\sqrt{x^2 + y^2 + z^2}} \right)
\end{equation}

\noindent where $x, y, z$ are as defined in Equation \ref{eq:coords}.

\begin{figure*}
    \centering
    \includegraphics[width = \textwidth]{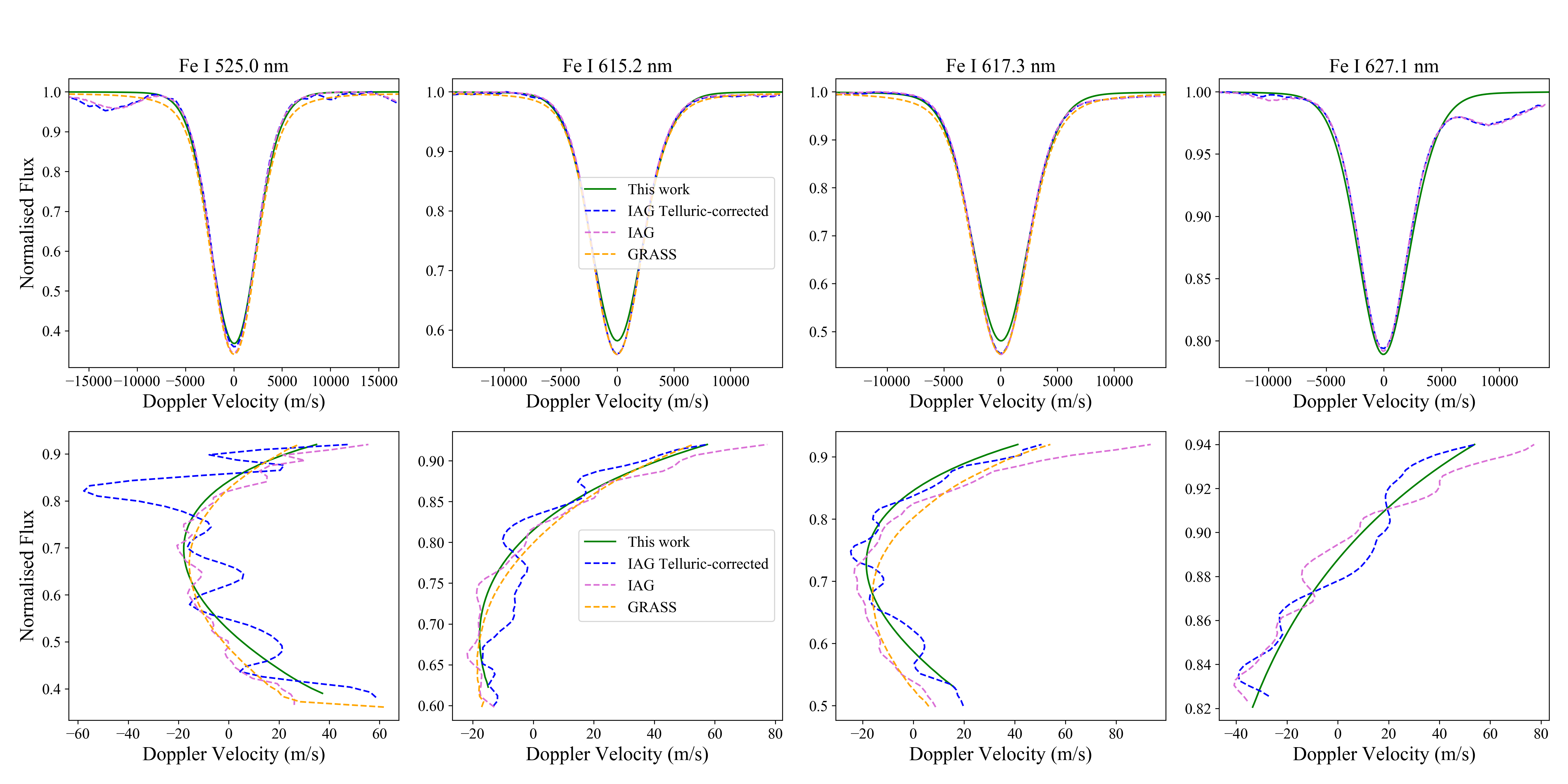}
    \caption{Comparisons of disk integrated results from this work with data from the IAG solar atlas \citep{Reiners2016}, the telluric-corrected version of the IAG solar atlas \citep{Baker2019} and results from GRASS \citep{grass2}. Note there is no GRASS result available for \ion{Fe}{I}~627.1nm, hence the omission. All data has been shifted so that the mean Doppler velocity of the bisector is 0 m/s. Wiggles in the bisectors of the telluric-corrected data are due to the wavelength grid being stored in single precision. Note that the flux axis for each line has a different scale.}
    \label{fig:iag}
\end{figure*}

\begin{figure}
    \centering
    \includegraphics[width = \columnwidth]{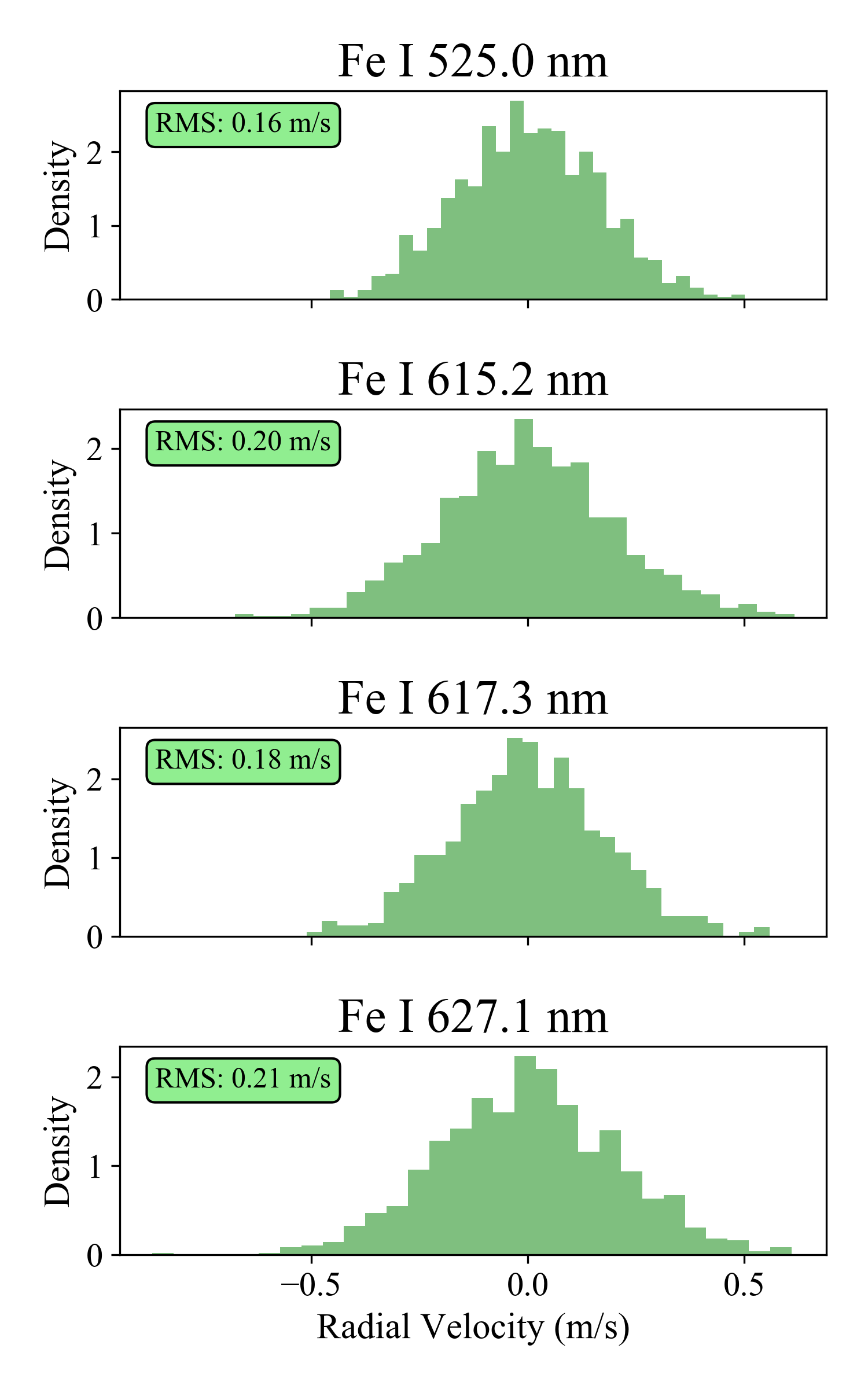}
    \caption{Granulation-induced velocity distributions across 1000 disk-integrated line profiles for each spectral line. The velocity values are calculated using a CCF with the mean profile. The RMS value of each distribution is printed in the green boxes.}
    \label{fig:rv_hist}
\end{figure}

\section{Disk integrated profiles with \texttt{DISCO}}
\label{sec:DI}

\texttt{DISCO} generates a profile $I(\theta)$ using Equation~\ref{eq:I} for each tile on the stellar grid. Note that $\epsilon_{1}$ and $\epsilon_{2}$ are random variables sampled from the generated distributions, and it is this that drives the granular variability. Our method does not impose filling factor correlations between neighbouring tiles. This reflects the fact that solar granules are small and short-lived features, with typical horizontal scales of order 1~Mm and lifetimes of only a few minutes. Granular evolution is dominated by local conditions, and there is no widely accepted mechanism that would maintain coherent behaviour between granules separated by several megametres. The tiles in our model span horizontal scales of 9~Mm, corresponding to approximately 5--10 granule diameters in the solar case. Observational studies of time-averaged solar velocity fields have identified persistent up- and down-flow structures on scales of only 2--4~Mm (e.g: \citealt{Baudin1997, Hirzberger1997}), which are significantly smaller than our tiles. Any such correlations are therefore expected to be contained within individual tiles rather than extending between neighbouring tiles. In addition, there is no clear observational evidence for a distinct mesogranular flow pattern \citep{Rincon2018}, supporting the assumption that tile-to-tile correlations can be neglected.

In each tile, the generated profile $I(\theta)$ is shifted in velocity space by the line of sight stellar rotational velocity calculated as in Section~\ref{ss:losvel} and multiplied by $A_p/\pi R_{\odot}^2$, where $A_P$ is the projected area from Section \ref{ss:proj_area}. The final disk integrated profile is then generated by summing over all of the visible tiles. 

Figure \ref{fig:iag} shows our disk-integrated results (the mean of 1000 instances) compared to observations from the Institut für Astrophysik, Göttingen (IAG) solar atlas \citep{Reiners2016}, as well as simulated results from \cite{grass2}. In this plot, all profiles have been shifted so that the mean Doppler velocity of the bisector is 0 m/s. In the profiles themselves, we find a strong match to observations other than our depth appears slightly shallower in all but \ion{Fe}{I} $627~\mathrm{nm}$. The depth mismatch is not introduced by the disk-integration procedure, as a similar, though much smaller, difference is already present when comparing disk-resolved profiles to observations. This suggests that the discrepancy originates in the local line synthesis, potentially due to uncertainties in the atomic data or missing physical effects such as magnetic fields, and is subsequently amplified through disk integration.

In the bisectors, we find strong matches to the telluric-corrected IAG solar atlas from \cite{Baker2019}. Note that the `wiggles' in the observational bisectors are due to the wavelength grid being saved in single precision and are not physical. As is evident from this plot, despite this wavelength region containing no strong telluric absorption lines, even tiny contaminants can cause velocity shifts of up to 10~m s$^{-1}$ at specific intensity levels. Our results are inherently telluric free as they are not influenced in any way by observational data. This agreement with the telluric-corrected dataset is an excellent indication that our interpolation across the stellar disk works as expected. 

By repeating calculations of disk integrated profiles over a number of iterations, we are able to ascertain the RV variability caused by the pure surface convection on a sun-like star. Our approach ensures that we have no contamination from p-modes, spots, faculae, super-granulation, meridional flows and tellurics, all of which will be present to some extent in observationally driven strategies. We also are free from concerns regarding instrumental precision and observing conditions. 

Figure \ref{fig:rv_hist} shows histograms of disk-integrated RVs for 1000 iterations of each line. The RVs here are calculated by cross-correlating each profile with the mean profile over all iterations. To extract the RV we fit a second order polynomial to the peak of the cross-correlation function. We find, as expected due to results reported in F25, that RV rms increases as lines get weaker. Our weakest line (\ion{Fe}{I} $627~\mathrm{nm}$) has an RV rms of $0.21~\mathrm{m~s^{-1}}$ whereas our strongest line (\ion{Fe}{I} $525~\mathrm{nm}$) has only $0.16~\mathrm{m~s^{-1}}$. All results are however lower than are reported in many empirically driven studies such as \cite{grass2}. We believe this is largely due to the true isolation of granulation impact. However, it is worth pointing out that our study ignores magnetic effects and the impact of this will require further investigation. 

Finally, we emphasize that the filling factors used in \texttt{DISCO} are drawn independently for each iteration, with no memory of previous realizations. As a result, the framework does not generate temporally connected time series, but instead produces statistically independent disk-integrated spectra. While this is sufficient for quantifying the intrinsic RV variability induced by granulation, it precludes the study of characteristic timescales or temporal correlations in the velocity signal. Extending this approach to produce physically motivated time series, for example by introducing random-walk or correlated stochastic evolution of the filling factors, represents a promising avenue for future work.

\section{Correcting for granulation-induced RV}
\label{sec:corr}

\begin{figure*}
    \centering
    \includegraphics[width = \textwidth]{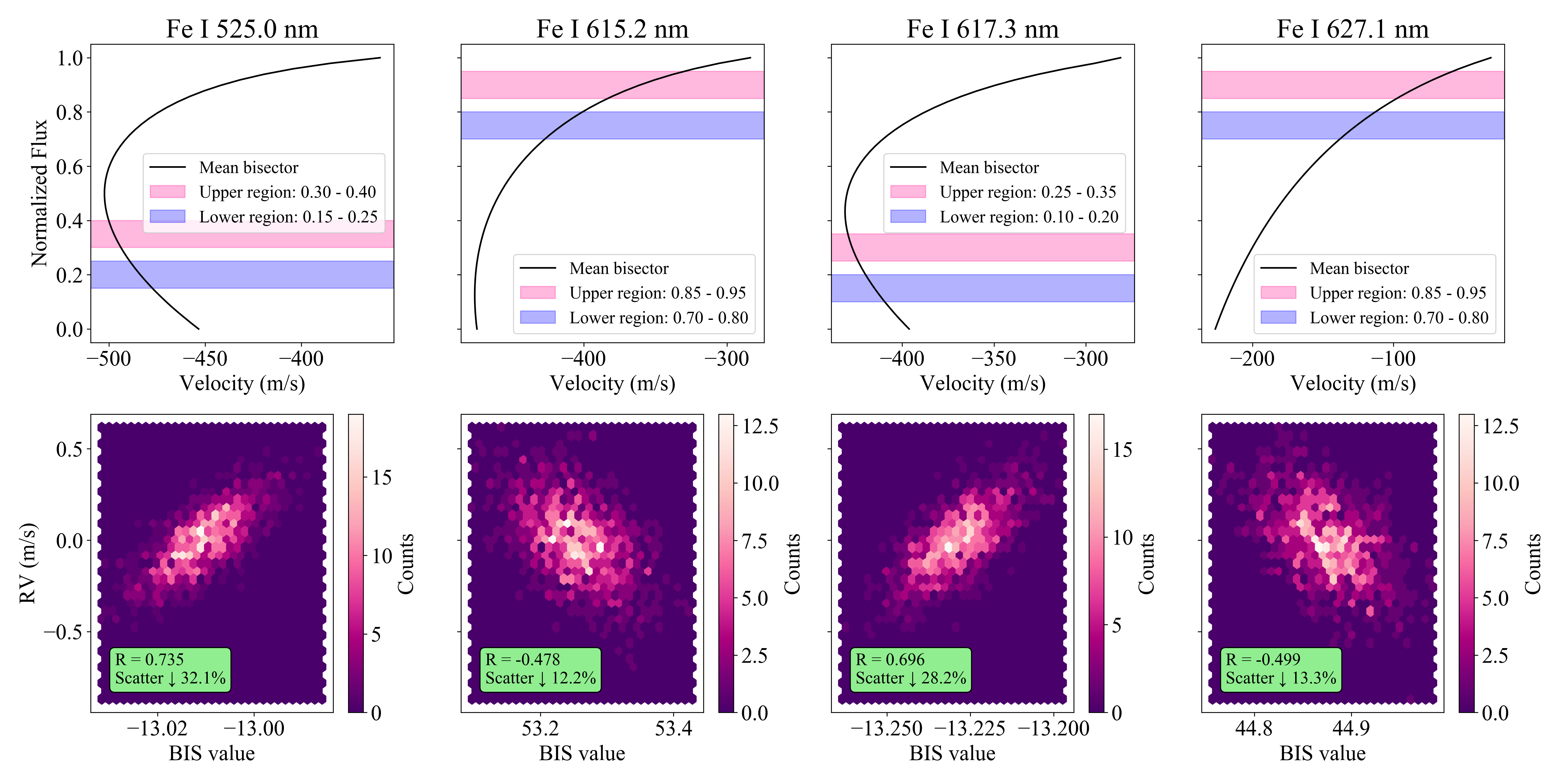}
    \caption{\textbf{Top panel:} Mean bisectors of each line. Shaded regions show the upper and lower regions that maximise the correlation between BIS values and granulation-induced velocity shifts. Normalised flux has been normalised according to the line depth. \textbf{Bottom panel:} Hexbin plots showing the relationship between BIS and RV for 1000 profiles for each spectral line. The pearsons R value of the correlation, along with the resulting decrease in RV scatter after correction is displayed in the textboxes. }
    \label{fig:bis_corr}
\end{figure*}

\begin{figure*}
    \centering
    \includegraphics[width = \textwidth]{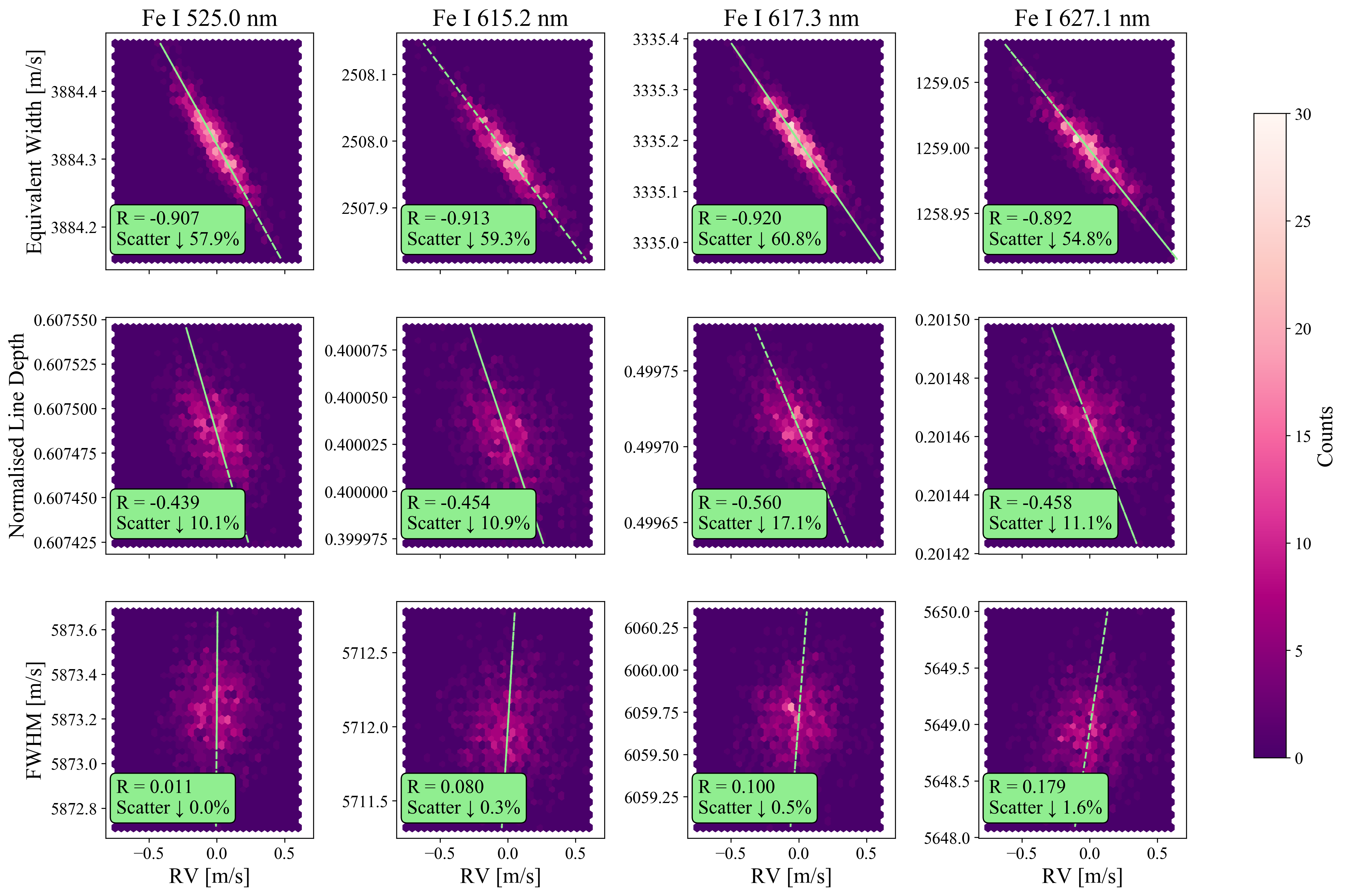}
    \caption{Relationships between granulation induced RV and some line shape diagnostics available through ground-based spectroscopy. The green line is a linear fit to the data and printed in the green box is the pearson R coefficient along with the RV rms reduction achievable using this result.}
    \label{fig:line_corr}
\end{figure*}

\begin{figure*}
    \centering
    \includegraphics[width = \textwidth]{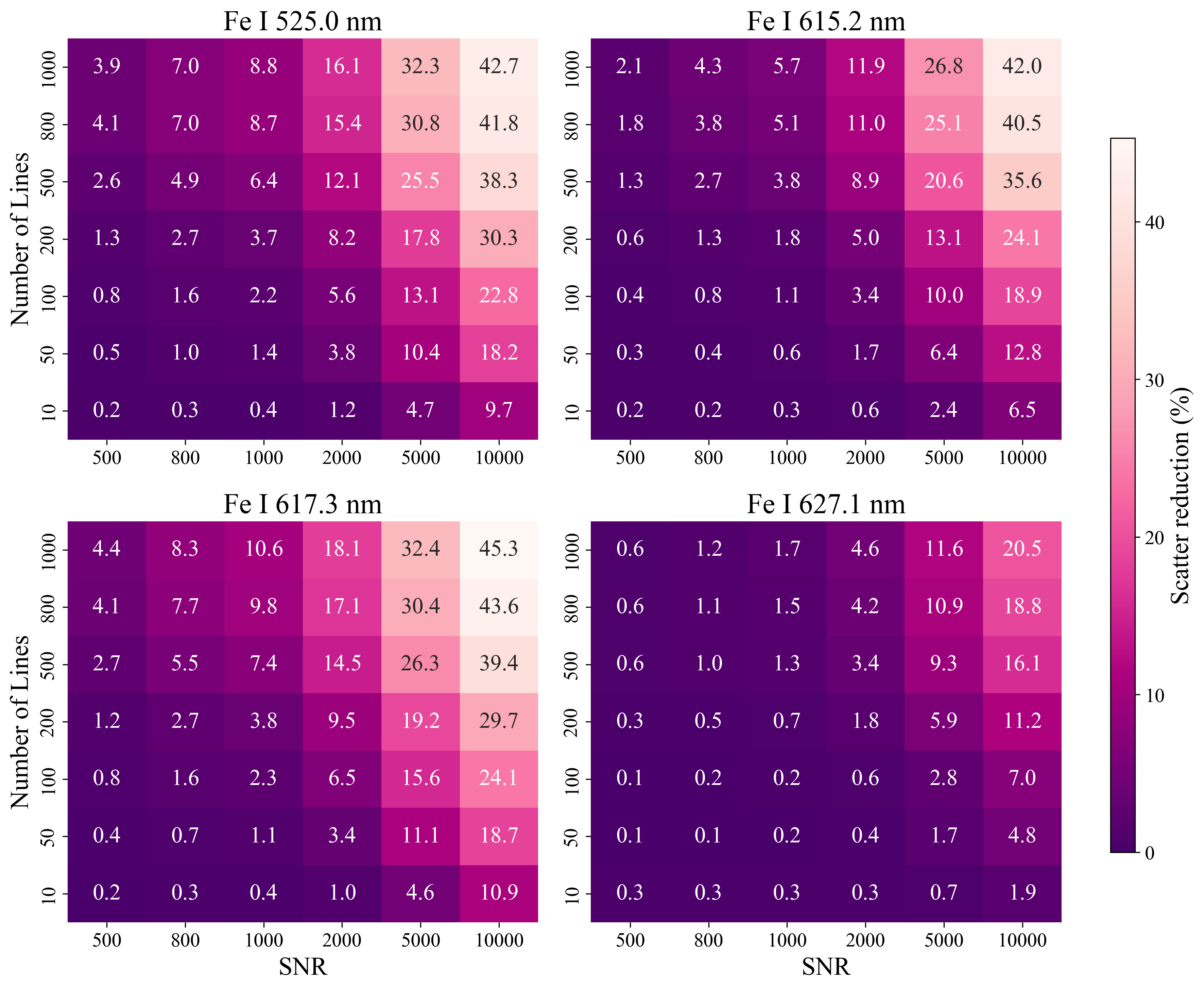}
    \caption{Heatmaps of RV rms scatter reduction for each line after a linear regression model trained on CCF shape diagnostics is applied. Results here have been averaged over 100 iterations. Varied parameters are the SNR per pixel and the number of lines included in the synthetic spectra.}
    \label{fig:heatmaps}
\end{figure*}



Whilst we have shown that granulation-induced RV rms may be lower than expected, it is still a contributing factor towards obscuring the $0.09~\mathrm{m~s^{-1}}$ RV signal imposed by an Earth twin. In this section we explore how effectively we can remove granulation-induced RV signals from a profile based only on the shape of the line. 


\subsection{Bisector asymmetry correlations}

In previous works such as \cite{cegla2019} and \cite{grass2}, correlations between bisector shape and granulation-induced RV have been used to mitigate velocity scatter. A key diagnostic of bisector asymmetry is the bisector inverse span (BIS), defined as the difference in the mean wavelength (or velocity) between an ``upper'' and ``lower'' region of the bisector. The conventional choice is 10--40\% and 55--90\% of the line depth. Both cited studies, however, report that the BIS--RV correlation for granulation can be significantly improved by fine-tuning these regions for each line profile. 

Following the optimisation strategy outlined in \cite{cegla2019}, we iteratively explore the parameter space from 5\% to 95\% of the line depth, subject to the constraints that each region spans at least 5 percentage points and that the two regions are separated by at least 5 percentage points. We choose our optimal regions based on maximising the absolute pearsons R value for the correlation between BIS and RV. Surprisingly, our optimal regions differ substantially from those reported by \cite{grass2} for the same lines. For bisectors with a pronounced ``C-shape,'' we find that both optimal regions lie below the C-bend, while for shallower lines without a C-shape both regions lie above 70\% of the line depth. 

In terms of scatter reduction, our BIS--RV correlations are broadly consistent with \cite{grass2}, yielding a $\sim$30\% reduction in RV rms for strong lines with C-shaped bisectors and $\sim$13\% for weaker lines. These improvements remain below those reported by \cite{cegla2019} for the \ion{Fe}{I}~630.2~nm line. A summary of our results is illustrated in Figure~\ref{fig:bis_corr}. We note that all of our optimised regions have a width of 10\% and are separated by 5\%. This pattern arises naturally from the optimisation and was not imposed as a constraint in the procedure.

We find that this method of reducing RV rms is highly sensitive to the precise definition of the bisector regions. Moreover, the signal-to-noise ratio of typical observational data is insufficient to determine BIS values accurately on a line-by-line basis. As noted by \cite{grass2}, the only viable observational application would involve grouping lines with similar sensitivity to bisector asymmetry and combining their signals. Although we achieve similar scatter reduction in our noise-free high resolution profiles to \cite{grass2}, the optimal regions we uncover cover much less of the bisectors shape than is traditional and therefore is much more impacted by noise and resolution.

We therefore view the BIS metric as a challenging diagnostic of granulation-induced variability. Its effectiveness is heavily dependent on the precise choice of regions and on the intrinsic shape of the spectral line, and it becomes increasingly unreliable at lower spectral resolution or in the presence of noise, since the interpolation required to construct the bisector amplifies measurement uncertainties. 

The BIS is, however, only one possible measure of line asymmetry. Other asymmetry diagnostics, such as those explored in \cite{cegla2019} have been shown to offer complementary and in some cases more robust sensitivity to granulation driven line distortions. A full exploration of the broader space of asymmetry based indicators is beyond the scope of this work, but represents an important direction for future studies. For the remainder of this paper, we restrict our analysis to simpler, widely used line-shape metrics.

\subsection{Fundamental line shape diagnostics}
\label{sec:LS_diag}

Ground-based spectroscopy does not retain information on the absolute continuum intensity. This limits the available line shape diagnostics to relative measures, such as equivalent width (EW), normalised line depth, and full width at half maximum (FWHM). We test the correlation of each of these diagnostics with the net convective blueshift across our line sample. 

We find that the strongest correlation is obtained with the equivalent width. The granulation induced RV scales tightly with EW, consistent with previous results \citep{Allende_Prieto1998, Ramirez2008, Reiners2016, cegla2019}. This relation is physically motivated: granules contribute line profiles that are both deeper and broader than those formed in intergranular lanes, mostly due to their strong temperature gradient, which leads to the line core forming at much cooler temperatures than the continuum. A larger filling factor of granules therefore yields both an increased EW and a stronger blueshift. In our dataset we find a Pearson correlation coefficient of $\sim -0.9$, corresponding to a scatter reduction of $\sim 50$–$60\%$ depending on the line. This highlights EW as the most effective full line shape diagnostic for mitigating convective blueshift variability.  
For the normalised line depth we find only a weak correlation with granulation induced RV, with Pearson coefficients of $\sim-0.4$, corresponding to a scatter reduction of around 10\%. The FWHM shows an even weaker correlation, with no significant improvement in scatter. See Figure \ref{fig:line_corr} for the relationships and linear fits for each line and diagnostic. We find the un-normalised line depth has a much stronger correlation, suggesting the EW variation is largely driven by continuum changes. High-precision photometry may therefore offer a promising complementary avenue for tracing granulation in future work.

We are able to further improve the scatter reduction by combining the information from all three diagnostics in a linear regression model. To simulate realistic observational conditions, we degrade our high-resolution, noise-free profiles to match the spectral resolution and sampling of the ESPRESSO spectrograph. Each simulated line profile is convolved with a Gaussian line-spread function (LSF) corresponding to the resolving power of the selected ESPRESSO mode (\(R = 70{,}000\), \(140{,}000\), or \(190{,}000\) for moderate, high and ultra-high resolution, respectively). Following convolution, the spectra are flux-conservingly rebinned onto a logarithmic wavelength gird, sampled at approximately four pixels per LSF FWHM. The rebinning procedure integrates overlapping wavelength intervals between the input and output grids, ensuring that total flux is conserved across bins. The resulting profiles thus reproduce the resolution and sampling characteristics expected from ESPRESSO observations under ideal, noise-free conditions.  The linear regression model is trained on 1000 instances and tested on a separate set of 1000 instances, with the resulting scatter reduction for each mode summarised in Table~\ref{tab:scat_red}. We find that spectral resolution has a negligible impact on these diagnostics, likely because the equivalent width is conserved through resolution changes.  

\begin{table}
\centering
\caption{Scatter reduction (\%) for each spectral line across ESPRESSO modes, after applying a linear regression model based on combined line shape diagnostics.}
\begin{tabular}{lcccccc}
\textbf{Line} & \multicolumn{3}{c}{\textbf{RV rms reduction (\%)}} \\
 & MR & HR & UHR \\
\hline
\ion{Fe}{I}~525.0~nm &  61.9 & 62.5 & 62.2 \\
\ion{Fe}{I}~615.2~nm & 71.5 & 72.0 & 71.9 \\
\ion{Fe}{I}~617.3~nm &  70.5 &  72.2 & 72.3 \\
\ion{Fe}{I}~627.1~nm & 63.2 & 68.0 & 68.1 \\
\hline
MR: R = 70,000 \\
HR: R = 140,000\\
UHR: R = 190,000
\end{tabular}
\label{tab:scat_red}
\end{table}

However, the results presented in this section correspond to ideal, noise-free conditions and serve only to demonstrate that the correlations exist in principle. In practice, observational noise will significantly hinder their recovery, and substantial work remains to assess their observability. The following section investigates how the model performs under realistic signal-to-noise ratios (SNR) and how the inclusion of additional spectral lines influences this behaviour.

\section{Effect of Photon Noise and Synthetic Spectra Generation}
\label{sec:noise}
  
The addition of photon noise to the line profiles reduces our ability to detect subtle correlations between line-shape diagnostics and granulation-induced radial velocities. In this study, we assume that the noise is dominated by photon noise, which scales with the signal intensity. To assess its impact, we inject Gaussian noise into the synthetic line profiles by perturbing each intensity value with \(\mathcal{N}(0, \sigma^2)\), where \(\sigma = I/\mathrm{SNR}\). Here, $I$ denotes the continuum-normalised intensity at each pixel, and SNR is the per-pixel signal-to-noise ratio. This represents a simplified treatment of realistic observational noise, as we assume a constant SNR across all wavelengths in our synthetic spectra.

We find that the correlations identified in Section~\ref{sec:LS_diag} rapidly diminish as noise increases, requiring an SNR greater than 100{,}000 for the relationships to remain discernible. It is important to note, however, that our analysis considers single-line diagnostics, whereas real observations typically encompass wide band spectra. The inclusion of multiple lines would introduce additional information and thus reduce the required per-pixel SNR.

To approximate a more realistic multi-line observation, we follow the procedure outlined in \cite{grass2} and construct synthetic spectra by replicating each single-line profile multiple times with wavelength offsets that include small random perturbations. Each replicated profile preserves the same granulation-induced Doppler shift and line-shape variations; only the photon-noise realisation differs between repetitions. This process extends each time-step spectrum into a broader wavelength range containing several copies of the same line, separated by a fixed amount but slightly perturbed to avoid aliasing. Alongside these extended spectra, we generate a corresponding template mask by placing Gaussian features at the rest wavelength of the line and then repeating them using the wavelength shifts applied in the synthetic spectra. This ensures the template is aligned with the artificial line pattern and provides a suitable reference for cross-correlation. The resulting synthetic spectra and template are then used to compute CCFs, allowing us to analyse how line-shape variations and noise influence the measured CCF and inferred velocity shifts. The more lines we include in our extended synthetic spectra, the more information is retained in our CCF, as noise contributions begin to average out. 

Varying both the SNR and the number of lines, we retrain and re-test the linear regression model introduced in Section~\ref{sec:LS_diag}, this time on the CCF rather than individual profiles. For consistency, we use the granulation-induced RVs derived from the noise-free spectra as the true targets, rather than recalculating RVs from the noisy spectra. Figure~\ref{fig:heatmaps} presents the resulting granulation induced RV scatter reduction achieved under different conditions. The outcomes are discouraging: even when combining 1000 strong lines at an SNR of 1000, the model achieves only a modest 8-11\% reduction in granulation induced RV scatter, with weaker lines yielding even poorer performance. These findings indicate that with these simple line shape diagnostics, granulation signatures are effectively obscured within the CCF shapes at realistic noise levels, rendering them undetectable with current instrumentation. This is in-keeping with results in \cite{Sulis2023}, who had difficulty isolating RV and CCF shape correlations in ESPRESSO spectroscopy for solar-like stars. Advances in either photon noise mitigation strategies, higher SNR observations, or the development of more sensitive diagnostics will be necessary to recover these subtle signals. 

The data presented in this work is stored in the public \texttt{DISCO} repository. All python scripts used to generate and analyse the profiles are also included. We welcome community suggestions or attempts to more effectively trace granulation signatures in the presence of photon noise.

\section{Summary and Conclusions}

This work builds on \cite{Frame2025} (F25), which demonstrated that parameterisation can be used to construct disk-resolved spectral absorption lines containing only the effects of granulation. In this paper, we have introduced an interpolation method that enables these profiles to be mapped onto a stellar grid with high accuracy.

We parameterise line profiles into three components reprensenting the intensity outputs of granular tops, outer granular regions and intergranular lanes.
We interpolate the time-averaged component profiles and sets of parameters from which we construct the filling-factor distributions for each component. The interpolation is trained on 16 limb angles, with 7 additional angles withheld for validation. As shown in Figure~\ref{fig:validation}, combining the interpolated profiles with the independently interpolated filling-factor distributions reproduces realistic profile samples at arbitrary limb angles, preserving the center-to-limb variations. This allows us to generate unlimited high-accuracy, time-varying profiles at minimal computational expense.

We tile a stellar grid with profiles generated in this manner and produce 1000 random instances of disk-integrated profiles for each of the four spectral absorption lines: \ion{Fe}{I} 525.0~nm, \ion{Fe}{I} 615.2~nm, \ion{Fe}{I} 617.3~nm, \ion{Fe}{I} 627.1~nm. The scripts to generate these profiles are available within the \texttt{DISCO} repository. These profiles are, by design, free from any contamination via p-modes, tellurics, meridional flows, instrumental noise and supergranulation; they contain only the effects of granulation. As a consequence of this, we are able to probe the velocity variations imposed by granulation, unencumbered by other effects. We find granulation induced RV rms values to be lower than is reported elsewhere, with values ranging from $0.16~\mathrm{m ~s^{-1}}$ for our strongest line (\ion{Fe}{I} 525.0~nm) to $0.21~\mathrm{m ~s^{-1}}$ for our weakest (\ion{Fe}{I} 627.1~nm).

We find that at individual disk positions, the optimal regions for BIS–RV correlations correspond to the traditional upper and lower parts of the bisector, yielding strong and consistent correlations. However, when integrating over the stellar disk, these clear relationships vanish. The optimal BIS regions in the disk-integrated profiles occupy only narrow portions of the bisector, making them highly sensitive to spectral resolution and noise. Although our results reproduce the magnitude of scatter reduction reported by previous studies for noise-free profiles, the correlations we obtain are less robust to noise. A more in-depth investigation of the effectiveness of alternative asymmetry measures is left for future work. We choose to focus on simpler metrics such as equivalent width, normalised line depth and full width at half maximum. 

Across our line sample, the equivalent width emerges as the most reliable diagnostic of granulation-induced velocity shifts, showing strong correlations with convective blueshift and reducing granulation induced RV scatter by up to 60\%. In contrast, normalised line depth and FWHM exhibit only weak or negligible correlations. Combining all three diagnostics in a linear regression model further improves performance, achieving up to 70\% scatter reduction in ideal, noise-free conditions, with negligible dependence on instrumental resolution. However, when realistic noise is introduced, these relationships rapidly degrade, remaining detectable only at unrealistically high signal-to-noise ratios (SNR $\sim$ 100,000) for single line analysis. Even when averaging over thousands of lines in synthetic multi-line spectra, scatter reductions remain below 10\% at SNRs typical of current high-resolution spectrographs. These results demonstrate that while these simple line-shape diagnostics contain valuable information on convective variability in principle, their practical utility is severely limited by observational noise and instrumental precision. 

To conclude, more powerful and noise-robust diagnostics will be essential for mitigating granulation signatures in observational data. The framework presented here enables the generation of an effectively unlimited training set of high-accuracy, disk-integrated line profiles that contain only granulation signals and come with known convective velocities. Such uncontaminated synthetic data provide a valuable resource for developing and benchmarking future mitigation strategies.

\section{Acknowledgments}

We would like to thank the anonymous referee for their insightful comments which improved this work.

This work acknowledges funding from a UKRI Future Leader Fellowship (grant numbers MR/S035214/1 and MR/Y011759/1). Computing facilities were provided by the Scientific Computing Research Technology Platform of the University of Warwick. GF acknowledges a Warwick prize scholarship (PhD) made possible thanks to a generous philanthropic donation. MLP was supported by the Flatiron Research Fellowship at the Flatiron Institute, a division of the Simons Foundation. CAW would like to acknowledge support from the UK Science and Technology Facilities Council (STFC, grant number ST/X00094X/1). AIS and VW acknowledge support from the European Research Council (ERC) under the European Union’s Horizon 2020 research and innovation program (grant no. 101118581). Computing facilities were provided by the Scientific Computing Research Technology Platform of the University of Warwick.

\section*{Data Availability}

All of the data required to generate the disk-integrated profiles studied in this work are stored in the public \texttt{DISCO} repository\footnote{\url{https://github.com/ginger-frame/DISCO}}, along with the python scripts to do so.



\bibliographystyle{mnras}
\bibliography{bib} 



\appendix

\section{Determination of Training Angles}
\label{sec:A1}

\begin{figure*}
    \centering
    \includegraphics[width = \textwidth]{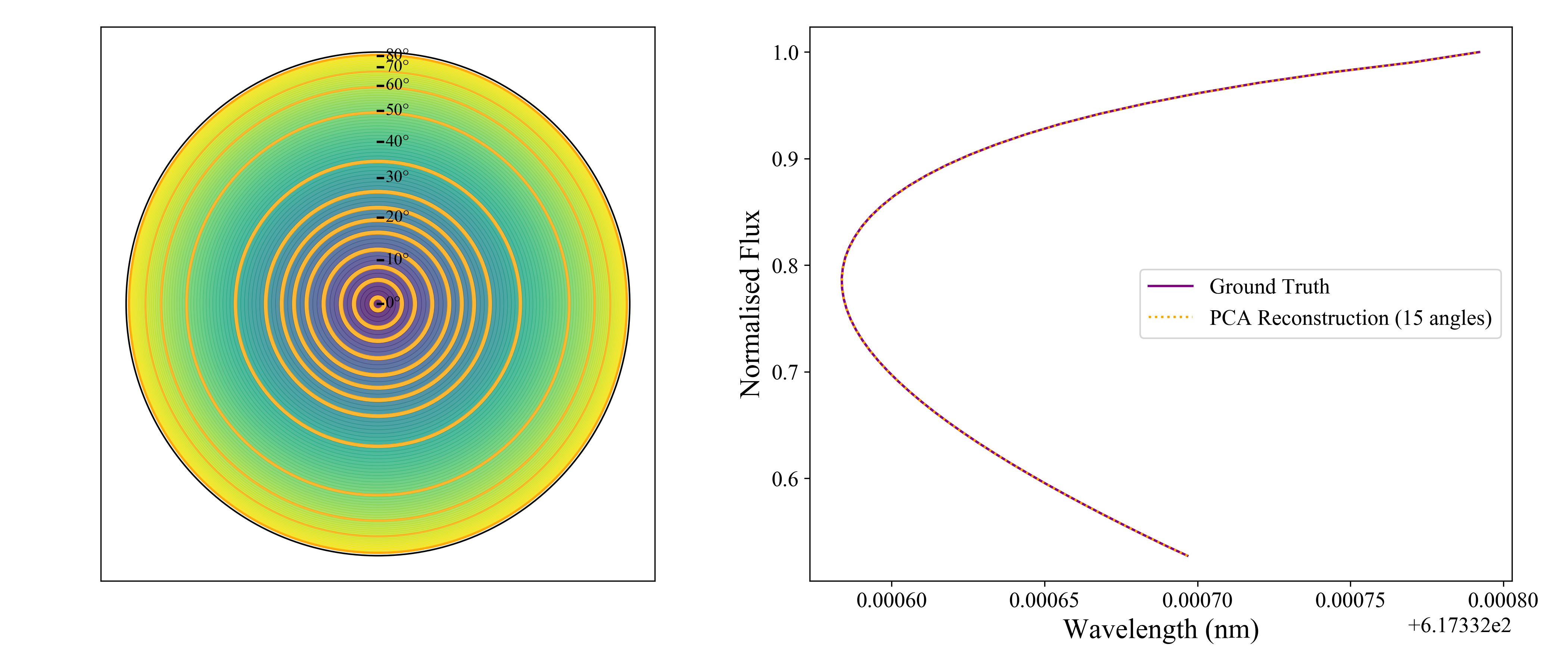}
    \caption{\textbf{Left plot:} Depiction of limb angle rings on the stellar grid. Colors represent the changing limb angle. Highlighted in orange are the 15 selected optimal training degrees for construction of the disk-integrated profile through PCA. \textbf{Right plot:} Bisector of disk integrated profiles using PCA trained on all angles with 14 components (Ground truth) versus the selected 15 angles with 7 components. The difference is indiscernible.}
    \label{fig:optimal_angles}
\end{figure*}

\subsection{Constructing a Ground Truth}

To establish a reference for evaluating our dimensionality-reduction approach, we selected four random snapshots from the time series and synthesised \ion{Fe}{I} 617~nm line profiles in $1^{\circ}$ increments from disk centre to a limb angle of $84^{\circ}$. The mean of the four snapshots at each angle was adopted as a proxy for the time-averaged profile.

To determine the optimal number of principal components to retain when applying PCA to reconstruct time-averaged profiles at arbitrary angles, we employed $k$-fold cross-validation with $k = 7$ \citep{stone1974}. In this procedure, the set of limb angles was partitioned into seven folds, and the PCA model was trained iteratively, each time withholding one fold as a validation set. For each iteration, reconstruction accuracy was quantified by computing the root mean squared error (RMSE) between the predicted validation profiles and their synthesised counterparts. This process was repeated for component numbers ranging from 3 to 20. The cross-validation results indicated that retaining 14 components yielded the lowest average RMSE, corresponding to an average reconstruction error of only 0.02\% in normalised flux. This choice thus provides a good balance between reducing dimensionality and avoiding overfitting.

Using this optimal PCA representation, we tiled the stellar grid and computed the disk-integrated profile following the procedure outlined in Section~\ref{sec:DI}, training the PCA model on all 84 angular positions with 14 retained components. For the full time series, however, it is computationally infeasible to synthesise profiles at every angular step. It is therefore necessary to identify a reduced set of training angles that can still capture the essential variability in the line profiles. The ground truth constructed above serves as a benchmark for this purpose, enabling a systematic comparison of different candidate training sets and the identification of an optimal subset of angles.

\subsection{Selection of Optimal Training Angles}

To identify the optimum reduced set of training angles, we adopted an iterative elimination strategy. Starting from the full set of 84 limb angles, we performed the disk integration repeatedly, each time withholding a single angle. In each case, the disk-integrated profile was reconstructed using PCA with seven retained components, and the reconstruction error was quantified via the RMSE relative to the ground truth profile defined in the previous section.

The reduction from fourteen to seven retained components was motivated by the smaller training sets encountered during this optimisation. Retaining too many components in this context increases the risk of overfitting, whereas seven components provided a robust representation of the profiles while still capturing the majority of the variance.

At each iteration, the angle whose removal caused the smallest increase in RMSE was permanently excluded from the training set. This process was repeated iteratively until only 15 angles remained. The resulting subset therefore represents the angular positions that contribute most significantly to preserving reconstruction accuracy while substantially reducing the computational expense of synthesising the full time series. Figure~\ref{fig:optimal_angles} illustrates the selected training angles and compares the resulting disk-integrated profiles against the ground truth. The relative RMSE between the ground truth profile and the reconstruction using only the 15 selected angles is approximately $0.0006\%$, indicating an excellent reconstruction.

Rather than repeating this computationally intensive procedure for each spectral line, we verified that the angle set derived for \ion{Fe}{I}~617~nm performs equally well for other lines. Computing the ground truth and reconstructing profiles using the 15 angles for additional lines yielded excellent agreement in all cases, with relative RMSE values remaining below $\sim0.001\%$.

In practice, interpolation must be performed on component profiles and granule filling factor parameters rather than a single time-averaged profile. To avoid any extrapolation issues at the disk centre, we therefore added $0^{\circ}$ to the training set, extending it to 16 angles. The final set of training angles is: 0, 2, 6, 9, 13, 17, 20, 23, 27, 35, 50, 60, 68, 81, 82, 83$^{\circ}$. For angles beyond 83$^{\circ}$ (up to 90$^{\circ}$), we treat them as 83$^{\circ}$ to avoid extrapolation. This is acceptable because the contribution from this extreme limb region is minimal due to its small projected area and strong limb darkening, and therefore has a negligible impact on the disk-integrated profile.

\section{Testing the recreation of filling factor distributions at arbitary limb angles}
\label{sec:B1}

\begin{figure*}
    \centering
    \includegraphics[width = \textwidth]{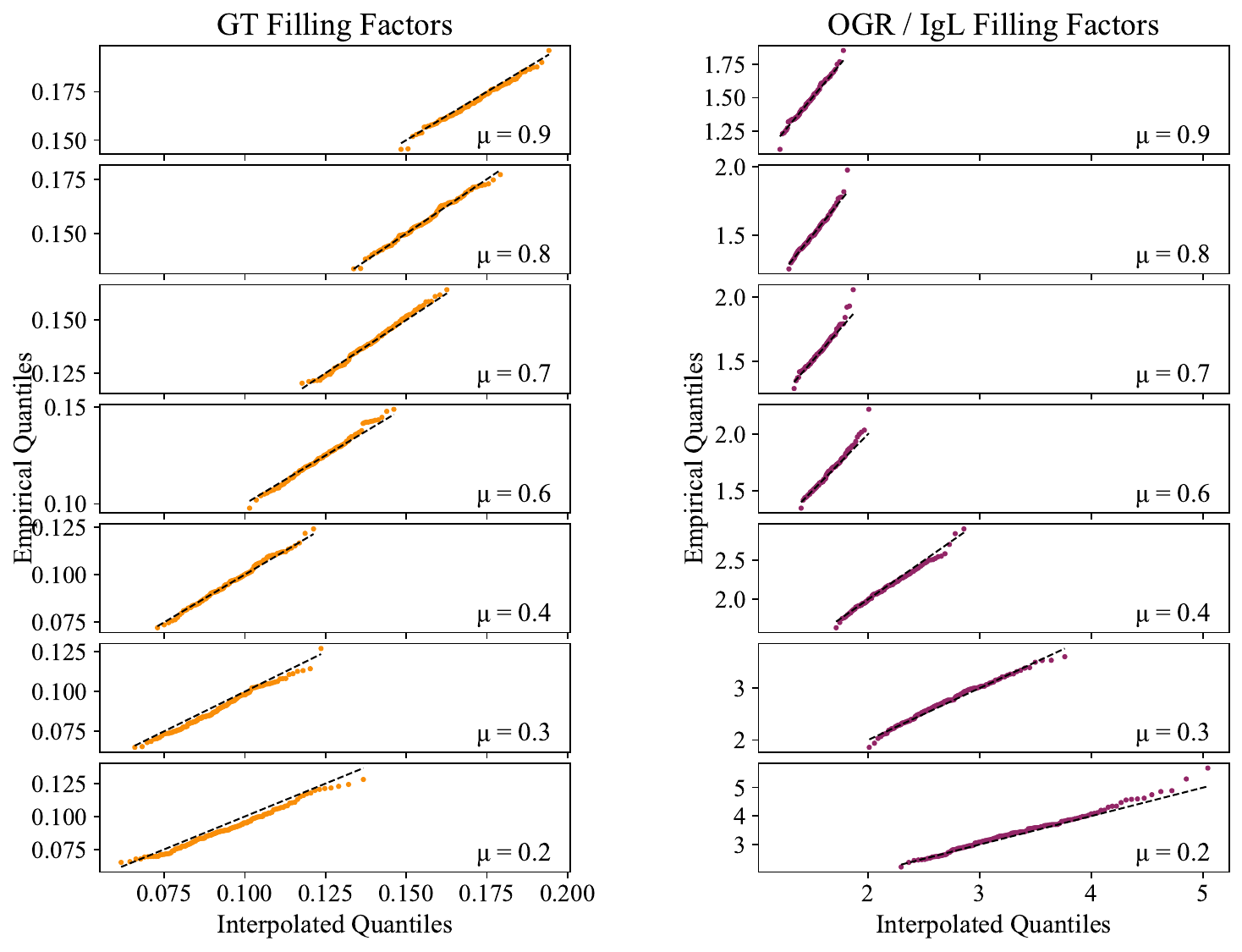}
    \caption{QQ plots showing the difference between interpolated and empirical quantiles for the GT and OGR/IgL filling factor distributions at different limb angles. The black dotted line indicates y=x, which is where the datapoints should lie if the interpolated and empirical distributions were exactly the same. The data shown here is from the validation set of limb angles and so has not been included in any training.}
    \label{fig:qq}
\end{figure*}

Figure~\ref{fig:qq} shows Q–Q plots comparing our interpolated skew-normal model to the empirical data at each validation angle. The black dashed line indicates the ideal $y=x$ reference; a perfect model fit would result in points closely following this line. Overall, the model shows good agreement, with a slight deviation toward the limb, as discussed in Section \ref{ss:ff}. Deviations at the distribution tails are likely due to under-sampling in the empirical dataset; 129 data points are likely insufficient to fully capture tail behavior.

\bsp	
\label{lastpage}
\end{document}